\documentclass{aastex}
\usepackage{emulateapj5}
\submitted{Received 2000 July 19; accepted 2000 October 6}
\shortauthors{SAIJO, SHIBATA, BAUMGARTE \& SHAPIRO}
\shorttitle{DYNAMICAL BAR INSTABILITY IN ROTATING STARS}
\begin{document}
%
\title
{Dynamical Bar Instability in Rotating Stars: Effect of General
Relativity}
%
\author
{Motoyuki Saijo,
Masaru Shibata\altaffilmark{1},
Thomas W. Baumgarte\altaffilmark{2}, and
Stuart L. Shapiro\altaffilmark{3}}
%
\affil
{ 
Department of Physics, University of Illinois at Urbana-Champaign,
Urbana, IL 61801-3080}
\altaffiltext{1}{Department of Earth and Space Science, 
Graduate School of Science, Osaka University,
Toyonaka, Osaka 560-0043, Japan}
\altaffiltext{2}{Fortner Fellow}
\altaffiltext{3}{Department of Astronomy and NCSA, University of Illinois at
Urbana-Champaign, Urbana, IL 61801}
%
%
%
\begin{abstract}
We study the dynamical stability against bar-mode deformation of
rapidly and differentially rotating stars in the first post-Newtonian
approximation of general relativity.  We vary the compaction of the
star $M/R$ (where $M$ is the gravitational mass and $R$ the equatorial
circumferential radius) between 0.01 and 0.05 to isolate the influence
of relativistic gravitation on the instability.  For compactions in
this moderate range, the critical value of $\beta \equiv T/W$ for the
onset of the dynamical instability (where $T$ is the rotational
kinetic energy and $W$ the gravitational binding energy) slightly
decreases from $\sim 0.26$ to $\sim 0.25$ with increasing compaction
for our choice of the differential rotational law. Combined with our
earlier findings based on simulations in full general relativity for
stars with higher compaction, we conclude that relativistic
gravitation {\em enhances} the dynamical bar-mode instability,
i.e.~the onset of instability sets in for smaller values of $\beta$ in
relativistic gravity than in Newtonian gravity.  We also find that
once a triaxial structure forms after the bar-mode perturbation
saturates in dynamically unstable stars, the triaxial shape is
maintained, at least for several rotational periods.  To check the
reliability of our numerical integrations, we verify that the general
relativistic Kelvin-Helmholtz circulation is well-conserved, in
addition to rest-mass energy, total mass-energy, linear and angular
momentum.  Conservation of circulation indicates that our code is not
seriously affected by numerical viscosity.  We determine the amplitude
and frequency of the quasi-periodic gravitational waves emitted during
the bar formation process using the quadrupole formula.
\end{abstract}

\section{Introduction}
\label{sec:intro}

Stars in nature are usually rotating and subject to nonaxisymmetric
rotational instabilities.  An exact treatment of these instabilities
exists only for incompressible equilibrium fluids in Newtonian gravity,
{\it e.g.} \citep{Chandra69,Tassoul,ST}.  For these configurations, 
global rotational instabilities arise from non-radial toroidal modes 
$e^{im\varphi}$
($m=\pm 1,\pm 2, \dots$) when $\beta\equiv T/W$ exceeds a certain
critical value. Here $\varphi$ is the azimuthal coordinate and $T$ and
$W$ are the rotational kinetic and gravitational potential binding
energies.  In the following we will focus on the $m=\pm 2$ bar-mode,
since it is the fastest growing mode when the rotation is sufficiently
rapid.

There exist two different mechanisms and corresponding timescales for
bar-mode instabilities.  Uniformly rotating, incompressible stars in
Newtonian theory are {\em secularly} unstable to bar-mode formation
when $\beta \geq \beta_{\rm sec} \simeq 0.14$.  However, this
instability can only grow in the presence of some dissipative
mechanism, like viscosity or gravitational radiation, and the growth
time is determined by the dissipative timescale, which is usually much
longer than the dynamical timescale of the system.  By contrast, a
{\em dynamical} instability to bar-mode formation sets in when $\beta
\geq \beta_{\rm dyn} \simeq 0.27$.  This instability is independent of
any dissipative mechanisms, and the growth time is the hydrodynamic
timescale of the system.

The secular instability in compressible stars, both uniformly and
differentially rotating, has been analyzed numerically within linear
perturbation theory, by means of a variational principle and trial
functions, by solving the eigenvalue problem, or by other approximate
means.  These techniques have been applied not only in Newtonian
theory \citep{LO67,OB73,FS78,IL90} but also in a
post-Newtonian (PN) framework 
(Cutler and Lindblom 1992; Shapiro and
Zane 1998 for incompressible stars) and in full general relativity
\citep{BFG96,BFG98,SF,YE}.  For relativistic stars, the
critical value of $\beta_{\rm sec}$ depends on the compaction $M/R$ of
the star (where $M$ is the gravitational mass and $R$ the
circumferential radius at the equator), on the rotational law and on
the dissipative mechanism.  The gravitational-radiation driven
instability occurs for smaller rotation rates, {\it i.e.}, for values
$\beta_{\rm sec} < 0.14$, in general relativity.  For extremely
compact stars (Stergioulas and Friedman 1998; Yoshida and Eriguchi
1999) or strongly differentially rotating stars (Imamura et al. 1995),
the critical value can be as small as $\beta_{\rm sec} < 0.1$. By
contrast, viscosity drives the instability to higher rotation rates
$\beta_{\rm sec} > 0.14$ as the configurations become more compact
\citep{BFG96,BFG98,SZ}.

Determining the onset of the dynamical bar-mode instability, as well
as the subsequent evolution of an unstable star, requires a fully
nonlinear hydrodynamic simulation.  Simulations performed in
Newtonian theory, {\it e.g.} \citep{TDM,DGTB,WT,HCS,SHC,HC,PDD,TIPD,NCT} 
have shown that
$\beta_{\rm dyn}$ depends only very weakly on the stiffness of the
equation of state.  Once a bar has developed, the formation of spiral
arms plays an important role in redistributing the angular momentum
and forming a core-halo structure. Recently, it has been shown that,
similar to the onset of secular instability, $\beta_{\rm dyn}$ can be
smaller for stars with a higher degree of differential rotation 
\citep{TH,PDD}.

To date, the dynamical bar-mode instability has been analyzed mainly
in Newtonian theory, since until quite recently a stable numerical
code capable of performing reliable hydrodynamic simulations in three
dimensions plus time in full general relativity has not existed.  Some
recent developments, however, have advanced the field significantly.
New formulations of the Einstein equation based on modifications of
the standard $(3+1)$ system of equations have resulted in codes which
have proven to be remarkably stable over many dynamical timescales,
{\it e.g.} \citep{SN,BS}.
In addition, gauge conditions which allow long-time stable evolution
for rotating, self-gravitating systems and are manageable
computationally have been developed, {\it e.g.} \citep{Shibata99b}.

The purpose of this paper is twofold.  We verify that the point of
onset of dynamical bar mode formation, as measured by $\beta$,
decreases with increasing compaction, and we furthermore show that
in unstable configurations, the bar persists for at least several
rotational periods.

In a previous paper \citet{SBS00} performed simulations of rapidly and
differentially rotating neutron stars in full general relativity.
They employed the relativistic code of \citet{Shibata99a} to study the
onset and growth of the dynamical bar-mode instability in relativistic
stars.  They found that for compact stars with $M/R \gtrsim 0.1$, the
onset of dynamical instability occurs at $\beta_{\rm dyn} \sim 0.24 -
0.25$, somewhat smaller than the Newtonian value of $\beta_{\rm dyn}
\sim 0.26$.  In principle, this reduction of $\beta_{\rm dyn}$ could
be caused by effects of either general relativity or differential
rotation.  In order to isolate the two effects one could study
sequences of varying compaction $M/R$ or sequences of varying degree
of differential rotation.  We choose the former in this paper and
focus on the {\em transition} to relativistic gravitation as the
compaction increases to moderate values.

In fully relativistic evolution codes the Courant condition
for the gravitational fields limits the size of the numerical
timesteps due to the speed of light\footnote{This limitation could
in principle be avoided by using an implicit finite difference
scheme.  However, such schemes are more complicated than explicit schemes, 
and have not yet been implemented for fully relativistic hydrodynamics in
three spatial dimensions}.  As a consequence, the ratio
between the dynamical timescale for the matter and the Courant
timestep for the fields scales approximately as $(M/R)^{-1/2}$, which
makes the calculation prohibitively slow for small compactions.  In
this paper, we therefore adopt the first order post-Newtonian (1PN)
code of \citet{SBS98} to study the effect of general relativity on
$\beta_{\rm dyn}$ for small and moderate compactions.  In this
formalism, the relativistic evolution equations for the gravitational
fields reduce to elliptic equations, which are solved on individual
timeslices and hence have no Courant condition. The numerical
timesteps are now limited by the Courant condition for the
hydrodynamical evolution equations alone and therefore scale with the
dynamical timescale.  We evolve models with compactions $M/R$ between
0.01 and 0.05 for several rotational periods to determine their
stability.  In cases in which a bar forms we also follow the nonlinear
growth and saturation of the instability.

Should the star retain its bar-like shape for many rotational periods,
then it would emit a quasi-periodic gravitational wave signal.  Such a
radiator would be potentially a very promising source for the
gravitational wave interferometers currently under construction.  This
prospect has prompted many researchers to study the evolution of the
bar-mode instability, most recently \citet{NCT} and \citet{Brown}.  They
have performed long-duration Newtonian simulations of dynamically
unstable stars that form a bar and then spiral arms.  These
configurations eject small amounts of mass and then settle down to
triaxial stars.  For the early stage of the evolution, the results of
these two groups are very similar, but for later times some
differences arise.  According to \citet{NCT}, the bar gradually decays
after a few rotational periods, probably due to numerical errors
associated with the unphysical motion of the center of mass, whereas
\citet{Brown} reports that the star remains bar-like for more than 10
rotational periods, and presumably until gravitational radiation
ultimately drives the decay of the bar.

Given these discrepancies, which are probably due to numerical
inaccuracies, it seems desirable to develop additional criteria to
verify the late-time reliability of numerical codes.  In particular,
it is possible that numerical viscosity associated with
finite-differencing could lead to an artificial decay of bar-modes.
In this paper we evaluate the conservation of circulation, which is
violated in the presence of viscosity (numerical or otherwise).  We
present a method for computing the circulation in general relativity
and demonstrate that, in our numerical code, the circulation is well
conserved.

This paper is organized as follows.  In Sec.~\ref{sec:model} we
present the basic equations of our 1PN formulation of general
relativity and describe our initial data in Sec.~\ref{sec:indata}.
We discuss our numerical results in Sec.~\ref{sec:stability}, focusing
on the dynamical stability of differentially rotating stars.  In
Sec.~\ref{sec:Circularity} we demonstrate the conservation of
circulation in our simulations and discuss the emitted gravitational
wave signal in Sec.~\ref{sec:GWaves}.  In Sec.~\ref{sec:Discussions}
we briefly summarize our findings.  Throughout this paper, we use the
geometrized units with $G=c=1$ and adopt Cartesian coordinates
$(x,y,z)$ with the coordinate time $t$.  Greek and Latin indices take
$(t, x, y, z)$ and $(x, y, z)$, respectively.

\section{Basic equations}
\label{sec:model}

In this section, we briefly review the 1PN formalism of \citet{SBS98}.
We solve the fully relativistic equations of hydrodynamics, but
neglect some higher-order PN terms in the Einstein field equations.

\subsection{The field equations}

Define the spatial projection tensor $\gamma^{\mu\nu} \equiv
g^{\mu\nu} + n^{\mu} n^{\nu}$, where $g^{\mu\nu}$ is the spacetime
metric, $n^{\mu} = (1/\alpha, -\beta^i/\alpha)$ the unit normal to a
spatial hypersurface, and where $\alpha$ and $\beta^i$ are the lapse
function and shift vector.  Within a 1PN approximation, the spatial
metric $g_{ij} = \gamma_{ij}$ may always be chosen to be conformally flat
\begin{eqnarray}
\gamma_{ij} = \psi^{4} \delta_{ij},
\end{eqnarray}
where $\psi$ is the conformal factor (see Chandrasekhar 1965;
Blanchet, Damour, and Sch\"afer 1989).  The spacetime line element
then reduces to
\begin{eqnarray}
ds^{2} &=&
( - \alpha^{2} + \beta_{k} \beta^{k} ) dt^{2} + 2 \beta_{i} dx^{i} dt 
\nonumber \\
&& + \psi^{4} \delta_{ij} dx^{i} dx^{j}.
\end{eqnarray}
We adopt maximal slicing, for which the trace of the extrinsic
curvature $K_{ij}$ vanishes,
\begin{equation}
K \equiv \gamma^{ij} K_{ij} = 0.
\end{equation}  
The 1PN field equations for the five unknowns $\psi$, $\alpha$ and
$\beta^i$ can then be derived conveniently from the (3+1) formalism.

Since the spatial metric is conformally flat, the transverse part of
its time derivative vanishes.  The transverse part of the evolution
equation of the spatial metric therefore relates the extrinsic
curvature to the shift vector,
\begin{equation}
2 \alpha \psi^{-4} K_{ij} =
\delta_{jl} \partial_{i} \beta^{l} + \delta_{il} \partial_{j} \beta^{l} - 
\frac{2}{3} \delta_{ij} \partial_{l} \beta^{l}.
\label{eqn:EvolutionMetric}
\end{equation}
Inserting Eq. (\ref{eqn:EvolutionMetric}) into the momentum constraint
equation then yields an equation for the shift vector $\beta^{i}$
\begin{eqnarray}
\label{eqn:MomentumC}
&&
\delta_{il} \triangle \beta^{l} + 
\frac{1}{3} \partial_{i} \partial_{l} \beta^{l} =16 \pi \alpha J_{i} +
\left(\partial_{j}\ln \left( \frac{\alpha}{\psi^{6}} \right)\right)
\nonumber \\
&& \times
\left( 
   \partial_{i} \beta^{j} + \delta_{il} \delta^{jk} \partial_{k} \beta^{l} -
  \frac{2}{3} \delta_{i}^{j} \partial_{l} \beta^{l} 
\right),
\end{eqnarray}
where $\Delta \equiv \delta^{ij}\partial_i\partial_j$ is the flat
space Laplacian and $J_{i} \equiv -n^{\mu} \gamma^{\nu}_{~i} T_{\mu
\nu}$ is the momentum density.  In the definition of $J_i$,
$T_{\mu\nu}$ is the stress energy tensor.

The conformal factor $\psi$ is determined from the Hamiltonian constraint
\begin{equation}
\triangle \psi =
- 2 \pi \psi^{5} \rho_{H} - \frac{1}{8} \psi^{5} K_{ij} K^{ij},
\label{eqn:HamiltonianC}
\end{equation}
where $\rho_{H} \equiv n^{\mu} n^{\nu} T_{\mu \nu}$ is the mass-energy
density measured by a normal observer.

Maximal slicing implies $\partial_t K = 0$, so that the trace of the 
evolution equation for the intrinsic curvature yields an equation for 
the lapse function $\alpha$,
\begin{equation}
\triangle (\alpha \psi) = 2 \pi \alpha \psi^{5} 
(\rho_{H} + 2 S) + \frac{7}{8} \alpha \psi^{5} K_{ij} K^{ij},
\label{eqn:evolution}
\end{equation}
where $S  = \gamma_{jk} T^{jk}$.  We also use Eq.
(\ref{eqn:HamiltonianC}) to derive this equation.

Equations~(\ref{eqn:MomentumC}) -- (\ref{eqn:evolution}) determine
the fully nonlinear relativistic metric for maximal slicing within the
conformal flatness approximation.  None of these equations involve
time derivatives, so that in a numerical finite difference
implementation the Courant condition is no longer coupled to the speed
of light.  Since the conformal flatness approximation introduces
errors at higher order than 1PN, it is reasonable to neglect terms in
these equations which are also higher than 1PN order.  In particular,
we discard $[ \partial_{j}[ \ln (\alpha / \psi^{6} ) ] ] ( \partial_{i}
\beta^{j} + \delta_{il} \delta^{jk} \partial_{k} \beta^{l} - 2
\delta_{i}^{j} \partial_{l} \beta^{l} / 3 ) $ in Eq.
(\ref{eqn:MomentumC}), $\psi^{5} K_{ij} K^{ij}/8$ in Eq.
(\ref{eqn:HamiltonianC}), and $7 \alpha \psi^{5} K_{ij} K^{ij} / 8$ in Eq.
(\ref{eqn:evolution}).  With these simplifications, the source terms on
the left hand sides of Eqs. (\ref{eqn:MomentumC}) --
(\ref{eqn:evolution}) become compact.  As a consequence, we can
solve these equations numerically very accurately with outer boundary
conditions set at fairly small distances outside the matter, and hence
on fairly small numerical grids.

The equation for the shift,
\begin{eqnarray}
\delta_{il} \triangle \beta^{l} + 
\frac{1}{3} \partial_{i} \partial_{l} \beta^{l} =16 \pi \alpha J_{i},
\label{shifteq}
\end{eqnarray}
can be further simplified by introducing a vector $B_{i}$ and a scalar 
$\chi$ according to
\begin{eqnarray}
\triangle B_{i} &=& 4 \pi \alpha J_{i},\\
\triangle \chi &=& - 4 \pi \alpha J_{i} x^{i}. 
\end{eqnarray}
The shift can then be computed from
\begin{equation}
\delta_{ij} \beta^{j} = 4 B_{i} - 
\frac{1}{2} [\partial_{i} \chi + \partial_{i} (B_{k} x^{k})],
\label{eqn:BetaDecomposition}
\end{equation}
and will automatically satisfy Eq.~(\ref{shifteq}).  The vector-type
Poisson equation (Eq. (\ref{shifteq})) for $\beta^i$ has hence been
reduced to four scalar-type Poisson equations for $B_i$ and $\chi$.

To summarize, we have reduced Einstein equations in a 1PN formalism
to six elliptic equations for the six variables ($\alpha \psi$, $\psi$,
$B_{i}$, $\chi$),
\begin{eqnarray}
\Delta (\alpha \psi) &=& 
2\pi\alpha \psi^5
(\rho_{H} + 2S) 
\equiv 4\pi S_{\alpha\psi}, 
\label{eq210} \\
\Delta \psi  &=& -2\pi \psi^{5} \rho_{H} \equiv 4\pi S_{\psi},
\label{eq211} \\
\Delta B_{i} &=& 4\pi\alpha J_{i},\label{eq212} \\
\Delta \chi  &=& -4\pi\alpha J_{i} x^{i}.\label{eq213}
\end{eqnarray}
These Poisson-type equations are solved imposing the following 
boundary condition at outer boundaries
\begin{eqnarray}
\alpha\psi &=&
1 - \frac{1}{r} \int S_{\alpha\psi} d^3x + O(r^{-3}),\\
\psi &= &
1 - \frac{1}{r} \int S_{\psi} d^3x + O(r^{-3}),\\
B_{x} &=& 
- \frac{x}{r^{3}} \int \alpha J_{x} x d^3x - 
\frac{y}{r^{3}} \int \alpha J_{x} y d^3x 
\nonumber \\
& & + O(r^{-4}), \\
B_{y} &=& 
- \frac{x}{r^{3}} \int \alpha J_{y} xd^3x - 
\frac{y}{r^{3}} \int \alpha J_{y} yd^3x  \nonumber \\
&& + O(r^{-4}) ,\\
B_{z} &=& 
- \frac{z}{r^{3}} \int \alpha J_{z} zd^3x + O(r^{-4}) ,\\
\chi &=& \frac{1}{r} \int \alpha J_{i} x^{i} d^3x + O(r^{-3}). 
\end{eqnarray}

\subsection{The matter equations}

For a perfect fluid, the energy momentum tensor takes the form 
\begin{equation}
T^{\mu \nu} = 
\rho \left( 1 + \varepsilon + \frac{P}{\rho} \right) u^{\mu} u^{\nu} +
Pg^{\mu\nu},
\end{equation}
where $\rho$ is the comoving rest-mass density, $\varepsilon$ the specific
internal energy, $P$ the pressure, and $u^{\mu}$ the four-velocity.

We adopt a $\Gamma$-law equation of state in the form
\begin{equation}
P = (\Gamma-1)\rho\varepsilon, \label{gammalaw1}
\end{equation}
where $\Gamma$ is the adiabatic index which we set to be $2$ in
this paper.

In the absence of thermal dissipation, Eq.~(\ref{gammalaw1}), together
with the first law of thermodynamics, implies a polytropic equation of
state
\begin{equation}
P = \kappa \rho^{1+1/n}, \label{gammalaw}
\end{equation}
where $n=1/(\Gamma-1)$ is the polytropic index and $\kappa$ is a
constant.  Thus, if the matter satisfies a polytropic equation of
state initially, the polytropic form of the equation of state is
preserved during the subsequent evolution in the absence of shocks.

From $\nabla_{\mu} T^{\mu\nu}=0$ together with the equation of state
(Eq. (\ref{gammalaw1})), we can derive the energy and Euler equations
according to
\begin{eqnarray}
&&
\frac{\partial e_{*}}{\partial t}+
\frac{\partial (e_{*} v^{j})}{\partial x^{j}} = 0
\label{eqn:Energy}
,\\
&&
\frac{\partial(\rho_{*} \tilde u_{i})}{\partial t}
+ \frac{\partial (\rho_* \tilde u_{i} v^{j})}{\partial x^{j}} 
=
- \alpha \psi^{6} P_{,i} 
- \rho_{*} \alpha \tilde u^{t} \alpha_{,i} 
\nonumber \\
&& \hspace{2cm}
+ \rho_{*} \tilde u_{j} \beta^{j}_{~,i}
+ \frac{2 \rho_{*} \tilde u_{k} \tilde u_{k}}{\psi^{5} \tilde u^{t}} 
\psi_{,i},  
\label{eqn:Euler}
\end{eqnarray}
where 
\begin{eqnarray}
e_{*} &=& (\rho \varepsilon)^{1/\Gamma} \alpha u^{t} \psi^{6},\\
v^{i} &=& {dx^i \over dt}=\frac{u^{i}}{u^{t}},\\
\rho_{*} &=& \rho \alpha u^{t} \psi^{6},\\
\tilde{u}^{t} &=& ( 1 + \Gamma \varepsilon ) u^{t},\\
\tilde{u}_{i} &=& ( 1 + \Gamma \varepsilon ) u_{i}.
\end{eqnarray}
Note that we treat the matter fully relativistically; the 1PN
approximation only enters through simplifications in the coupling to
the gravitational fields.  Note also that we do not need to include an
artificial viscosity, since we do not encounter any shocks in the
simulations in this paper.  As a consequence we also do not need to
solve the continuity equation
\begin{equation}
\frac{\partial \rho_{*}}{\partial t}
+\frac{\partial (\rho_{*} v^{i})}{\partial x^{i}} = 0,
\label{eqn:BConservation}
\end{equation}
since in the absence of shocks it is equivalent to
Eq.~(\ref{eqn:Energy}).

The gravitational mass $M$, {\it e.g.} \citep{BY}, rest mass
$M_{0}$, proper mass $M_{p}$,  angular momentum $J$, kinetic energy
$T$, and gravitational binding energy $W$ of a rotating star can be
computed  from 
\begin{eqnarray}
M &=& - \frac{1}{2 \pi} \oint_{\infty} \nabla^{i} \psi dS_{i}
\nonumber \\
&=&
\int \biggl[ ( \rho + \rho \varepsilon + P ) (\alpha u^{t})^{2} - P \biggl]
\psi^{5} d^{3}x
,\\
M_{0} &\equiv&
\int \rho u^{t} \sqrt{-g} d^{3} x
= \int \rho_{*} d^{3} x
,\\
M_{p} &=&
\int \rho u^{t} ( 1 + \epsilon ) \sqrt{-g} d^{3} x
= \int \rho_{*} ( 1 + \varepsilon ) d^{3} x
,\\
J &=&
\int T^{t}_{\phi} \sqrt{-g} d^{3} x 
= \int (x J_{y} - y J_{x}) \psi^{6} d^{3} x
,\\
T &=&
\frac{1}{2} \int \Omega T^{t}_{\phi} \sqrt{-g} d^{3} x 
\nonumber \\
&=&
\frac{1}{2} \int \Omega (x J_{y} - y J_{x}) \psi^{6} d^{3} x
,\\
W &=&
M_{p} + T - M.
\end{eqnarray}
We also define the quadrupole moments $I_{ij}$ as
\begin{equation}
I_{ij} = \int \rho_{*} x^{i} x^{j} d^{3}x
,
\end{equation}
and the nondimensional, scale-invariant ratio $\beta \equiv T/W$, 
which is very useful to characterize the dynamical stability against
bar-mode deformation.

Since we use a polytropic equation of state, it is convenient to
rescale all quantities with respect to $\kappa$.  Since $\kappa^{n/2}$
has dimensions of length, we introduce the following nondimensional
variables
\begin{equation}
\begin{array}{c c c}
\bar{t} = \kappa^{-n/2} t
, &
\bar{\tau} = \kappa^{-n/2} \tau
, &
\bar{x} = \kappa^{-n/2} x
,\\
\bar{y} = \kappa^{-n/2} y
, &
\bar{z} = \kappa^{-n/2} z
, &
\bar{\Omega} = \kappa^{n/2} \Omega
,\\
\bar{M} = \kappa^{-n/2} M
, &
\bar{R} = \kappa^{-n/2} R
, &
\bar{C} = \kappa^{-n/2} C
\end{array}
\end{equation}
where $\tau$ is the proper time, $\Omega$ the angular velocity of the
star, and $C$ circulation (see Sec. 5 for a definition).  We
also define the central rotation period as $P_{c}=2\pi/\Omega_{0}$. 
Henceforth, we adopt nondimensional quantities, but omit the bars for
convenience (equivalently, we set $\kappa = 1$).

\section{Initial Data}
\label{sec:indata}

To prepare initial data, we construct axisymmetric rotating stars in
equilibrium and slightly perturb them.  The equilibrium configurations
are obtained by solving the equation of hydrostatic equilibrium
together with the field equations for metric ({\it i.e.},
Eqs. (\ref{eq210}) -- (\ref{eq213})).  For stationary solutions, the 1PN
Euler equation can be integrated to yield the 1PN Bernoulli equation
\begin{eqnarray}
&& \ln (1+\Gamma\varepsilon)+{1 \over 2}
\ln[\alpha^2-\psi^4\varpi^2(\Omega+\beta^{\varphi})^2]\nonumber \\
&& +\int u^t u_{\varphi} d\Omega={\rm const},
\label{hydrost}
\end{eqnarray}
where $\varpi=\sqrt{x^2+y^2}$ and
$\beta^{\varphi}=(x\beta^y-y\beta^x)/\varpi^2$.

Following previous studies 
\citep{KEH89a,KEH89b,CST92,CST94,BGSM, SBGH,SBS00},
we adopt the differential rotation law 
\begin{equation}
F(\Omega) \equiv u^{t} u_{\varphi} = A^{2} (\Omega_{0} - \Omega),
\end{equation}
where $A$ is a constant with dimension of length and $\Omega_{0}$ is
the angular velocity on the rotational axis.  With this choice, the
hydrostatic equation (Eq.~(\ref{hydrost})) can be integrated
analytically.  In the Newtonian limit ( $u^{t} \rightarrow 1$ and
$u_{\varphi} \rightarrow \varpi^2 \Omega$) the corresponding
rotational profile reduces to
\begin{equation}
\Omega = \frac{A^{2} \Omega_{0}}{\varpi^{2} + A^{2}}.
\label{NewOmega}
\end{equation}
This equation implies that $A$ determines the length scale over which
$\Omega$ changes, so that a smaller value of $A$ yields a larger
degree of differential rotation.  As in \citet{SBS00}, we choose 
$A=r_{e}$, where $r_{e}$ is the equatorial
coordinate radius of the star, which corresponds to a moderate degree
of differential rotation.  This choice allows us to compare directly
with the results of \citet{SBS00} and to
focus on the effects of general relativity.  Note that only uniformly
rotating stars with sufficiently stiff equations of state ($n \lesssim
0.5$) become dynamically unstable to bar formation before reaching the
mass shedding limit, so that the effects of general relativity on
$\beta_{\rm dyn}$ have to be studied for differentially rotating stars
for typical configurations.

\begin{table*}[b]
\begin{center}
\tablenum{1}
\label{tbl:initial}
\centerline{\sc Table 1}
\centerline{\sc Differentially rotating stars in equilibrium.}
\vskip 6pt
\begin{tabular}{c c c c c c c c c c c}
\hline
\hline
Model & 
$r_p/r_e$ &
$\bar{\rho}_{\rm max}$ &
$\bar{P}_{\rm c}$ & 
$\bar{P}_{\rm e}$ &
$\bar{M}$ & 
$\bar{M}_{0}$ & 
$\bar{J}$ & 
$T/W$ & 
$R/M$ & 
stability\\ 
\hline
I (a) & $0.250$ & $0.0117$ & 
$37.40$ & $83.52$ &
$0.113$ & $0.123$ & $0.0264$ & 
$0.265$ & $20.01$ & unstable
\\
I (b) & $0.275$ & $0.0125$ & 
$36.03$ & $80.50$ &
$0.110$ & $0.120$ & $0.0245$ &
$0.259$ & $19.94$ & unstable
\\
I (c) & $0.300$ & $0.0135$ & 
$34.97$ & $78.27$ &
$0.107$ & $0.116$ & $0.0221$ & 
$0.249$ & $20.00$ & unstable
\\
I (d) & $0.325$ & $0.0148$ & 
$33.96$ & $75.97$ &
$0.109$ & $0.103$ & $0.0198$ & 
$0.238$ & $20.01$ & stable
\\
\hline
II (a) & $0.250$ & $0.00655$ & 
$52.39$ & $111.6$ &
$0.0711$ & $0.0746$ & $0.0128$ & 
$0.271$ & $34.34$ & unstable
\\
II (b) & $0.275$ & $0.00753$ & 
$50.21$ & $107.1$ &
$0.0700$ & $0.0735$ & $0.0120$ & 
$0.263$ & $34.04$ & unstable
\\
II (c) & $0.300$ & $0.00821$ & 
$47.19$ & $100.9$ &
$0.0709$ & $0.0746$ & $0.0116$ & 
$0.252$ & $32.37$ & stable
\\
II (d) & $0.325$ & $0.00901$ & 
$46.19$ & $98.48$ &
$0.0675$ & $0.0710$ & $0.0102$ &
$0.240$ & $32.75$ & stable
\\
\hline
III (a) & $0.250$ & $0.00359$ & 
$72.64$ & $150.6$ &
$0.0417$ & $0.0428$ & $0.00566$ &
$0.273$ & $61.36$ & unstable
\\
III (b) & $0.275$ & $0.00381$ & 
$70.78$ & $146.7$ &
$0.0401$ & $0.0412$ & $0.00511$ &
$0.265$ & $62.46$ & unstable
\\
III (c) & $0.300$ & $0.00430$ & 
$67.95$ & $140.4$ &
$0.0393$ & $0.0404$ & $0.00470$ &
$0.254$ & $61.73$ & stable
\\
III (d) & $0.325$ & $0.00480$ & 
$66.15$ & $137.1$ &
$0.0377$ & $0.0388$ & $0.00417$ &
$0.241$ & $61.76$ & stable
\\
\hline
IV (a) & $0.250$ & $0.00218$ & 
$94.58$ & $193.1$ &
$0.0262$ & $0.0266$ & $0.00279$ &
$0.275$ & $100.1$ & unstable
\\
IV (b) & $0.275$ & $0.00236$ & 
$91.39$ & $186.6$ &
$0.0256$ & $0.0260$ & $0.00258$ & 
$0.267$ & $100.2$ & unstable
\\
IV (c) & $0.300$ & $0.00265$ & 
$88.21$ & $180.7$ &
$0.0248$ & $0.0251$ & $0.00232$ &
$0.255$ & $100.4$ & stable
\\
IV (d) & $0.325$ & $0.00296$ & 
$85.81$ & $174.6$ &
$0.0238$ & $0.0242$ & $0.00207$ &
$0.242$ & $100.1$ & stable
\\
\hline
\hline
\end{tabular}
\vskip 12pt
\begin{minipage}{15cm}
{
Note. ---
$\bar{\rho}_{\rm max}$: maximum rest-mass density;
$\bar{P}_{\rm c}$: rotational period along the rotational axis;
$\bar{P}_{\rm e}$: rotational period at the equator;
$\bar{M}$: gravitational mass;
$\bar{M}_{0}$: rest mass;
$\bar{J}$: angular momentum;
$T$: rotational kinetic energy;
$W$: gravitational binding energy;
$R$: equatorial circumferential radius.
}
\end{minipage}
\end{center}
\end{table*}

We prepare initial conditions for various values of the compaction
$M/R$ and rotation rate.  The latter is conveniently parameterized by
the deformation $r_p/r_e$, where $r_p$ is the polar coordinate
radius.  Since we adopt a 1PN approximation all results are correct
only up to the linear order in $GM/c^2R$ (or $(v/c)^2$ where $v$ is a
typical speed), implying that we can expect reliable results only for
moderate compactions $M/R \ll O(1)$.  We therefore restrict our analysis
to compactions in the range between 0.01 and 0.05.  In \citet{SBS00},
we found that fully relativistic stars with $M/R \sim 0.10 - 0.15$ and
$A=r_e$ become dynamically unstable against bar-mode formation
when $\beta \gtrsim 0.25$.  Taking this result as a guide, we
prepare rotating stars with $\beta \sim 0.25$, for which the
corresponding deformation $r_p/r_e$ takes values $\sim 0.25-0.33$.  
We label different initial data models with a Roman number and a Latin
letter, {\it e.g.}, Model II (c), where the Roman number labels the 
compaction, and the Latin letter the deformation, hence $\beta$.
We tabulate the physical parameters of our initial value models in 
Table \ref{tbl:initial}.

In order trigger bar-mode formation in unstable models, we slightly
perturb the density of the equilibrium models according to
\begin{equation}
\rho = \rho^{\rm (equilibrium)} 
\left( 1 + 0.1 \times \frac{x^{2}-y^{2}}{r_{e}^{2}} \right).
\label{equ:perturbation}
\end{equation}
This perturbation affects only the $I_{xx}$ and $I_{yy}$ components of
the quadrupole moment, which change by approximately $1\%$.  This
perturbation can therefore be considered linear initially.

Both for the construction of initial data and their subsequent
evolution, we assume planar symmetry across the equator, and solve the
equations on a uniform grid of typical size $101 \times 101 \times
51$.  In the axisymmetric initial configuration, the star's major and
minor axes are covered by $40$ and $10-13$ grid points.  Dynamically
unstable stars with high values of $\beta$ form bars and eject mass.
To avoid significant mass outflow across the outer boundaries during
the simulation, we use a larger grid of $141\times 141 \times 71$
gridpoints for these models.  We terminate any simulation when 1\% of
the total rest mass has been ejected from the numerical grid or the
evolution time reaches around $8 P_{c}$.  Our longest runs took about
18000 time steps, and consumed about 80 CPU hours on a VX/4R
vector-parallel computer at the National Astronomical Observatory of
Japan.

\section{Dynamical stability of differentially rotating stars}
\label{sec:stability}

We evaluate the stability of the perturbed rotating stars by
monitoring the distortion parameter
\begin{eqnarray}
\eta \equiv \frac{I_{xx}-I_{yy}}{I_{xx}+I_{yy}},
\end{eqnarray}
which is a measure of the magnitude of the bar-mode perturbation.
In Fig.~\ref{fig:Moment}, we show $\eta$ as a function of time for all
our models.  When the star is dynamically unstable, $\eta$ grows
exponentially up to a saturation level at which $\eta = O(1)$.  Once
the perturbation has saturated, the maximum value of $\eta$ remains
nearly constant on dynamical timescales implying that the bar
structure persists.  For stable stars, the maximum value of $\eta \ll
1$ remains approximately constant throughout the evolution.

The early exponential growth in unstable stars can be seen more
clearly in Fig.~\ref{fig:DetStability}, where we plot $|\eta|$ as a
function of time on a logarithmic scale.  We can determine the growth
time $\tau_g$ and the oscillation period $\tau_o$ by fitting to a
function
\begin{equation} \label{equ:etafit}
\eta = \eta_0 10^{t/\tau_g} \cos(2\pi t/\tau_o+\varphi_0),
\label{etaeta}
\end{equation}
where $\eta_0$ and $\varphi_0$ are constants.  We tabulate $\tau_g$
and $\tau_o$ for the unstable stars in Table~2.  Interestingly,
$\tau_o$ depends only very weakly on $M/R$ and $\beta$, which agrees
with the fully relativistic findings of \cite{SBS00}.

\vskip 18pt
\begin{minipage}{8cm} 
\begin{center}
\tablenum{2}
\label{tbl:stability1}
\centerline{\sc Table 2}
\centerline{\sc $\tau_o$ and $\tau_g$ in the early 
stage of bar formation.}
\vskip 6pt
\begin{tabular}{ccc}
\hline
\hline
Model &
$\tau_o/P_{c}$ &
$\tau_g/P_{c}$\\ 
\hline
I (a) & 1.30 & 1.54 
\\
I (b) & 1.31 & 2.17
\\
I (c) &  1.29 & 3.52
\\
\hline
II (a) & 1.30 & 1.57
\\
II (b) & 1.31 & 2.30
\\
\hline
III (a) & 1.31 & 1.60
\\
III (b) & 1.31 & 2.45
\\
\hline
IV (a) & 1.30 & 1.61
\\
IV (b) & 1.30 & 2.63
\\
\hline
\hline
\end{tabular}
\end{center}
\vskip -6pt
{Note. --- $\tau_o$: oscillation period of bar-mode perturbation. 
$\tau_g$: growth time of bar-mode perturbation (see 
Eq. (\ref{equ:etafit})). }
\end{minipage}
\vskip 18pt

Stable stars may show signs of an exponential growth very early on,
but their distortion parameter always remains very small $\eta \ll
O(1)$.  As a criterion for stability, we therefore check whether
$\eta$ follows an exponential growth as in Eq.~(\ref{etaeta}) up to
large values $\eta=O(1)$.  Judging from this criterion, Models I (a),
(b), (c), II (a), (b), III (a), (b), and IV (a) and (b) are
dynamically unstable.  We summarize these results in
Fig.~\ref{fig:crit}, where we denote stable and unstable models in a
$M/R$ versus $\beta$ plane.  The critical value $\beta_{\rm dyn}$
approaches $\sim 0.26$ in the Newtonian limit $M/R \rightarrow 0$, and
decreases to $\sim 0.25$ for $M/R=0.05$.  Our Newtonian value differs
slightly from the result for uniformly rotating, incompressible stars
($\beta = 0.27$), which is most likely due to the differential
rotation in our models.  This result confirms the fully relativistic 
results of \cite{SBS00}, which
we have also included in Fig.~\ref{fig:crit}.  There, we showed that
$\beta_{\rm dyn}$ decreases to $\lesssim 0.24$ for large compactions,
$M/R \gtrsim 0.1$.

In Figs. \ref{fig:D20} -- \ref{fig:D100}, we show contours of the 
density $\rho_*$ in the equatorial plane for the final stages of the 
dynamical evolution.  These plots clearly exhibit a triaxial
structure for the unstable models, while for stable models the
density distribution hardly changes during the evolution.

Once the bars in dynamically unstable models have saturated, they
persist with very little change for several rotation periods (see,
{\it e.g.}, Models I (c) and IV (b) in Fig. \ref{fig:Moment}).  In
realistic systems, gravitational radiation reaction would provide a
dissipation mechanism which would cause a decay of the bars.  Using
the quadrupole formula
\begin{equation}
\dot{E} \sim \frac{32}{5} \Omega_0^{6} (I_{xx} - I_{yy})^{2} 
\sim M^{2} R^{4} \Omega_0^{6} \eta^{2}
,
\end{equation}
(see, {\it e.g.}, Lai and Shapiro 1995), we can estimate the ratio
between the radiation reaction timescale $\tau_{\rm GW}$ and the
central rotation period as
\begin{equation}
\frac{\tau_{\rm GW}}{P_{c}} \sim \frac{1}{\eta^{2}}
\left( \frac{R}{M} \right)^{5/2} 
\left( \frac{M/R^3}{\Omega_0^2} \right)^{3/2}
.\label{reaction}
\end{equation}
Eq.~(\ref{reaction}) implies that as long as the star is not extremely
compact $M/R = O(1)$, the timescale of the radiation reaction is much
longer than the dynamical timescale of the system, even if its
rotation rate is near break-up, $\Omega_0^2 \sim M/R^3$, and the star
is highly deformed, $\eta=O(1)$. In our simulation we choose $M/R
\lesssim 0.05$ so that $\tau_{\rm GW}/P_c \gg 1$.  Gravitational
radiation reaction, which is not present at 1PN order, is therefore
irrelevant for the calculations in this paper.

\section{Conservation of Circulation}
\label{sec:Circularity}

Considerable effort recently has gone into determining whether or
not a bar, once it has saturated, persists for many rotational
periods, or whether it decays very soon (see, {\it e.g.}, New,
Centrella, and Tohline 2000; Brown 2000).  Differences in the
results are probably due to numerical errors, and possibly caused by
the presence of numerical viscosity associated with the finite
differencing.  In this section we describe how the conservation of
circulation in general relativity can be used to check for the
presence of numerical viscosity and to establish the reliability
of long-time simulations.

According to the Kelvin-Helmholtz theorem, the relativistic
circulation
\begin{equation}
{\cal C}(c) = \oint_{c} h u_{\mu} \lambda^{\mu} d\sigma, 
\end{equation}
is conserved in isentropic flow along an arbitrary closed curve $c$
(see Carter 1979; Landau and Lifshitz 1982). Here,
$h=1+\varepsilon+P/\rho$ is the specific enthalpy, $\sigma$ is a
Lagrange parameter which labels points on the curve $c$, and
$\lambda^{\mu}$ is the tangent vector to the curve $c$ [i.e.,
$\lambda^{\mu}=(\partial/\partial
\sigma)^{\mu}$].  Conservation of $C$ can be verified by computing
\begin{eqnarray}
\frac{d}{d\tau} C(c) &=&
\oint_{c} d\sigma u^{\nu} \nabla_{\nu} (h u_{\mu}
\lambda^{\mu}) 
\nonumber \\
&=&
\oint_{c} d\sigma 
\left[ 
  \lambda^{\mu} u^{\nu} \nabla_{\nu} (h u_{\mu}) + 
  (h u_{\mu}) u^{\nu} \nabla_{\nu} \lambda^{\mu} \right] 
\nonumber \\
&=&
\oint_{c} d\sigma [\lambda^{\mu} u^{\nu} \nabla_{\nu} (h u_{\mu}) +
h u_{\mu} \lambda^{\nu} \nabla_{\nu} u^{\mu}] 
\nonumber \\
&=&
- \oint_{c}  d\sigma \lambda^{\mu} \nabla_{\mu} h 
\nonumber \\
&=&
0.
\end{eqnarray}
Here, to derive the third line from the second line, we use 
the fact that $u^{\mu} = (\partial/\partial
\tau)^{\mu}$ and $\lambda^{\mu}$ are coordinate basis
vectors, and thus commute according to
\begin{equation}
u^{\mu} \nabla_{\mu} \lambda^{\nu} = 
\lambda^{\mu} \nabla_{\mu} u^{\nu}. 
\end{equation}
We also have used $u_{\mu}u^{\mu}=-1$ and the Euler equation
\begin{equation}
u^{\lambda} \nabla_{\lambda} (h u_{\mu}) = - \nabla_{\mu} h,
\end{equation}
to obtain the fourth line.  Note that it is the derivative of $C(c)$ with
respect to the {\em proper} time $\tau$ that vanishes, so that the
circulation has to be evaluated on hypersurfaces of constant proper
time as opposed to constant coordinate time.

\begin{table*}[b]
\begin{center}
\tablenum{3}
\label{tbl:circulation}
\centerline{\sc Table 3}
\centerline{\sc Relative error of the circulation.}
\vskip 6pt
\begin{tabular}{c c c c c}
\hline
\hline
Model &
$r/r_{e}$ &
Initial value &
Final value &
Relative error
\\ 
\hline
I (b) & 0.500 & 1.198 & 1.203 & 0.4\%
\\
I (b) & 0.625 & 1.651 & 1.632 & 1.1\%
\\
I (b) &  0.750 & 2.053 & 1.899 & 7.5\%
\\
\hline
I (c) & 0.500 & 1.173 & 1.187 & 1.2\%
\\
I (c) & 0.625 & 1.603 & 1.600 & 0.2\%
\\
I (c) & 0.750 & 1.989 & 1.992 & 0.2\%
\\
I (c) & 0.875 & 2.330 & 2.356 & 1.1\%
\\
\hline
I (d) & 0.500 & 1.129 & 1.142 & 1.2\%
\\
I (d) & 0.625 & 1.535 & 1.536 & 0.1\%
\\
I (d) & 0.750 & 1.903 & 1.910 & 0.4\%
\\
I (d) & 0.875 & 2.228 & 2.257 & 1.3\%
\\
\hline
\hline
\end{tabular}
\vskip 12pt
\begin{minipage}{10.5cm}
{Note. --- Conservation of circulation for Models I (a), (c) and (d).
We set each of the final circulation value at the end point
in Fig. \ref{fig:cc}.}
\end{minipage}
\end{center}
\end{table*}

We check the conservation of circulation for three cases, namely the
unstable Models I (b) and (c), and the stable Model I (d). 
For Model I (b) we evaluate the circulation along
three closed curves, and for Models I (c) and (d) for four.  Each
loop is located in the equatorial plane, and is initially aligned with a
constant density contour.  For Model I (b) we choose loops
which intersect $x/r_{e}$ = 0.5, 0.625, and 0.75 on the $x$ 
axis, and for Models I (c) and (d) an additional loop which 
intersects $x/r_{e}$ =0.875 and $y=0$.  
We follow the evolution of each loop with the help of 
Lagrangian tracers, whose trajectories are computed from
\begin{equation} 
\frac{dx^i}{dt} = v^i, 
\hspace{5mm} 
\frac{d\tau}{dt} = \frac{1}{u^t}.
\label{particle}
\end{equation} 
The number of test particles representing each loop is $80-140$
depending on its initial location.  We use a first order numerical scheme
to integrate Eqs.~(\ref{particle}) forward in time, once the (Eulerian)
hydrodynamic flow has been determined.  We evaluate the
circulation along a loop $c$ at a proper time $\tau$ by interpolating
the hydrodynamic variables in both space and time to the current
location of the Lagrangian tracers.

Fig. \ref{fig:cc} shows that the circulation is well conserved in all
three Models, indicating that numerical viscosity only has very small
effects in our code.  In Fig. \ref{fig:ptc}, we also show the location of
the loops (Lagrangian particles) at the final time steps. As expected,
the curves for the dynamically unstable stars (Models I (b) and (c))
become deformed, while the curves for the dynamically stable star
(Model I (d)) remain close to spherical.  The outermost loop in Fig.
\ref{fig:cc} (a) seems to indicate a small violation of the conservation
of circulation.  However, this loop is close to the surface of a star
which forms spiral arms, and is hence strongly deformed (see Fig.
\ref{fig:ptc}).  The representation of this loop with Lagrangian tracers
was therefore not sufficient to accurately evaluate its circulation.
We tabulate the relative errors in the circulation in Table
\ref{tbl:circulation}.  Except for the outermost loop in  Model I(b), 
the circulation is conserved up to $\sim 1$\% in our simulations.

We would like to emphasize that circulation is conserved in relativity
even in the presence of gravitational radiation (as long as the fluid
flow is isentropic and no shocks form).  As a consequence,
conservation of circulation can be used as a very strong code test in
fully relativistic simulations as well as Newtonian or post-Newtonian
simulations which include gravitational radiation reaction terms.

\section{Gravitational Waves}
\label{sec:GWaves}
We compute approximate gravitational waveforms by evaluating the
quadrupole formula, neglecting all PN corrections. 
In the radiation zone, gravitational waves can be described by a
transverse-traceless, perturbed metric $h_{ij}^{TT}$ with respect to 
a flat spacetime. In the quadrupole formula, $h_{ij}^{TT}$ 
can be expressed as \citep{MTW}
\begin{equation}
h_{ij}^{TT}= 
\frac{2}{r} \frac{d^{2}}{d t^{2}} 
\left( I_{ij}^{TT} - \frac{1}{3} \delta_{ij} I_{kk}^{TT} \right), 
\label{eqn:wave1}
\end{equation}
where $r$ is the distance to the source, and
$TT$ denotes the transverse-traceless projection 
\begin{equation}
I_{ij}^{TT} = P_{i}^{~a} P_j^{~b} I_{ab} - \frac{1}{2} P_{ij} P^{ab} I_{ab},
\end{equation}
with
\begin{equation}
P_{i}^{~j} = \delta_{i}^{~j} - \hat{n}_{i} \hat{n}^{j}, 
\hspace{5mm} \hat{n}^{i} = x^{i}/r.
\end{equation}
Choosing the direction of the wave propagation to be along the $z$
axis, the two polarization modes of gravitational waves can be
determined from
\begin{equation}
h_{+} \equiv \frac{1}{2} (h_{xx}^{TT} - h_{yy}^{TT}), 
\hspace{5mm} 
h_{\times} \equiv h_{xy}^{TT}.
\end{equation}
Note that this quantity contains second time derivatives of $I_{ij}^{TT}$,
which are difficult to evaluate numerically.   We therefore rewrite Eq.
(\ref{eqn:wave1}) as 
\begin{equation}
h_{ij}^{TT} =
\frac{2}{r} \frac{d}{dt} 
\left( \dot{I}_{ij}^{TT} - \frac{1}{3} \delta_{ij} \dot{I}_{kk}^{TT} \right),
\end{equation}
and use the continuity equation (Eq. (\ref{eqn:BConservation})) to 
eliminate the time derivatives in $\dot{I}_{ij}^{TT}$
\begin{equation}
\dot{I}_{ij} = \int (\rho_{*} v^{i} x^{j} + \rho_{*} x^{i} v^{j}) d^{3}x
,
\end{equation}
{\it e.g.} \citep{F89}.  We are then left with having to compute only
a first time derivative numerically, which can be done much more
accurately.  For observers along the $z$-axis, we find
\begin{eqnarray}
\frac{r h_{+}}{M} &=& 
\frac{1}{2M} \frac{d}{d t} (\dot{I}_{xx} - \dot{I}_{yy}),
\\
\frac{r h_{\times}}{M} &=& 
\frac{1}{M} \frac{d}{d t} \dot{I}_{xy}
.
\end{eqnarray}

In Fig. \ref{fig:gw}, we plot waveforms for Models I (b), (c), and (d).
For the unstable Models I (b) and (c), we find, as expected, a
quasi-periodic oscillation with growing amplitude during the early bar
formation.  For the stable Model I (d), we find a periodic waveform with
approximately constant amplitude.

In all three models, the frequency of the periodic oscillations
is approximately
\begin{equation}
f \sim {1 \over 1.3P_c} \simeq 770{\rm Hz} 
\biggl( {1~{\rm msec} \over P_c} \biggr). 
\label{frequency}
\end{equation}
Once a bar forms and reaches saturation in unstable stars, the
oscillation period reduces to slightly smaller values ($\simeq 1.18
P_{c}$ for Model I (b) and $ \sim 1.27 P_{c}$ for Model I (c)), and
accordingly the frequency shifts to slightly larger values.  This
shift in the frequency is caused by the significant deformation of the
stars by the bars.

The gravitational wave amplitude in Model I (b) is approximately 
\begin{equation}
h_{\rm GW}   \simeq   4.8
\times 10^{-23} \left(  \frac{M}{M_{\odot}} \right) 
\left(\frac{10{\rm Mpc}}{r} \right) 
\left( \frac{r h_{+,\times}/M}{0.01}\right).
\end{equation}
Both the frequency and
gravitational wave amplitude are consistent with the fully
relativistic results in \citet{SBS00}.

A detection of these signals by kilometer size laser-interferometric
gravitational wave detectors like LIGO, {\it e.g.} \citep{KIP} might
be feasible, because the wave form is quasi-periodic, which
significantly increases its effective amplitude (see Lai and Shapiro
1995).  Also, radiation reaction may gradually shift the frequency to
smaller values, and hence into a regime where LIGO is more sensitive.
The effect of radiation reaction on the evolution of bar modes is not
very well understood, though, except in incompressible stars,
{\it e.g.} \citep{Chandra70,LS} and should be the subject of future 
studies.

\section{Summary}
\label{sec:Discussions}
We perform 1PN simulations of rapidly and differentially rotating
stars to investigate general relativistic effects on the dynamical
bar-mode instability for small compactions $M/R \leq 0.05$.  By
combining these PN results with the fully relativistic simulations of
\cite{SBS00} for configurations of higher compaction, we conclude that
the critical value of $\beta = \beta_{\rm dyn}$ decreases with
increasing $M/R$.  Thus, relativistic gravitation enhances the
bar-mode instability.

We also describe how conservation of circulation can be used to check
by how much a code is affected by numerical visocity.  In the presence
of significant numerical viscosity, long-time evolution calculations
become very unreliable and may lead, for example, to erroneous
evolution of a saturated bar.  We show that in our calculations the
circulation is well conserved, implying that our code is at most very
weakly affected by numerical viscosity.  We present a method for
computing the circulation which can be applied in Newtonian,
post-Newtonian and fully relativistic calculations.

\acknowledgments
This work was supported by JSPS Grant-in-Aid (No. 5689), NSF Grants
AST 96-18524 and PHY 99-02833 and NASA Grant NAG5-7152 at the
University of Illinois at Urbana-Champaign.  M.~Shibata gratefully
acknowledges support by JSPS (Fellowships for Research Abroad) and the
hospitality of the Department of Physics at UIUC; T.~W.~Baumgarte is
pleased to acknowledge support through a Fortner Fellowship.
Numerical computations were performed on the VX/4R machines in the
Astronomical Data Analysis Center of National Astronomical Observatory
of Japan, and on the FUJITSU-VX/S vector computer at Media Network
Center, Waseda University.


\clearpage
\figcaption[Fig01.eps]{
Deformation parameter $\eta$ as a function of
$t/P_c$ for our sixteen different models (see Table I).  Solid, 
dashed, dash-dotted and dotted lines denote Models (a),
(b), (c) and (d), respectively.  We terminate the simulations after $t
\sim 8 P_{c}$, or when 1\% of the total rest mass has escaped from the
computational grid (Models I (a), (b), II (a) (b), III (a), and IV
(a)).
\label{fig:Moment}}

\figcaption[Fig02.eps]{   
Same as Fig.~\ref{fig:Moment}, but we show $|\eta|$ on a logarithmic
scale for $t/P_c \leq 4$.
\label{fig:DetStability}}

\figcaption[Fig03.eps]{
Summary of our dynamical stability analysis.  All our models are
plotted in a $\beta$ versus
$M/R$ plane, with stable stars denoted by a circle and unstable
stars by a triangle.  The solid circles and triangles are the models
studied in this paper; the open circles and triangles are the models
explored in full GR by \citet{SBS00}.  We conclude that the critical
value of $\beta = \beta_{\rm dyn}$ slightly decreases with increasing
compaction $M/R$.  This trend is emphasized by the dotted line, which
shows $\beta_{\rm dyn}$ as a function of $M/R$ approximately.
\label{fig:crit}}

\figcaption[Fig04.eps]{
Final density contours for $\rho_*$ in the
equatorial plane for Models I.  The contour lines denote densities
$\rho_{*}=1.3~i \times 10^{-3}$ ($i=1, \ldots, 15$) and at times
(a) $t/P_{c}=2.72$, (b) $t/P_{c}=3.66$, (c) $t/P_{c}=7.77$,
and (d) $t/P_{c}=8.16$.
\label{fig:D20}}

\figcaption[Fig05.eps]{
Same as Fig. \ref{fig:D20} for Models II.
The contour lines denote densities $\rho_{*}=6.0~i \times
10^{-4}$ ($i=1, \ldots, 15$) at times (a) $t/P_{c}=2.80$, (b)
$t/P_{c}=4.27$, (c) $t/P_{c}=8.38$, and (d) $t/P_{c}=8.25$.
\label{fig:D30}}

\figcaption[Fig06.eps]{
Same as Fig. \ref{fig:D20} for Models III.
The contour lines denote densities $\rho_{*}=3.1~i \times
10^{-4}$ ($i=1, \ldots, 15$), at times (a) $t/P_{c}=2.91$, (b)
$t/P_{c}=4.25$, (c) $t/P_{c}=8.04$, and (d) $t/P_{c}=7.93$.
\label{fig:D60}}

\figcaption[Fig07.eps]{
Same as Fig. \ref{fig:D20} for Models IV.  The
contour lines denote densities $\rho_{*}=1.7~i \times 10^{-4}$
($i=1, \ldots, 14$), at times (a) $t/P_{c}=2.77$, (b) $t/P_{c}=7.72$,
(c) $t/P_{c}=9.86$, and (d) $t/P_{c}=9.73$.
\label{fig:D100}}

\figcaption[Fig08.eps]{
Circulation as a function of proper time $\tau$ for various
loops in Models I (b), (c) and (d).  The loops have an initial
radius of $r/r_e = 0.5$ (solid line), 0.625 (dashed line), 0.75
(dash-dotted line) and 0.875 (dotted line).
\label{fig:cc}}

\figcaption[Fig09.eps]{
Location of Lagrangian test particles at the end
of the simulations for Model I (b) (at $\tau/P_c = 7.37$, $7.26$ listed
from the outer loop), Model I (c) (at $\tau/P_c = 11.06$, $10.95$,
$10.84$, $10.75$) and Model I (d) (at $\tau/P_c = 7.67$, $7.59$, $7.51$,
$7.45$).  Each loop is plotted for a constant value of
$\tau$, but different loops correspond to slightly different values of
$\tau$.  We also include density contours for $\rho_{*} = 10^{-4}
\rho_{* \rm max}$ (solid lines), which nearly coincide with the
surface of the star.  The long dashed, dashed, dash-dotted, dotted
lines denote the locations of the test particles that are initially
located along loops which intersect $x/r_{e}=0.875$, $0.750$, $0.625$,
and $0.500$ on the $x$ axis.
\label{fig:ptc}}

\figcaption[Fig10.eps]{
Gravitational waveforms $r h_+ /M$ (solid lines) and $r h_{\times}/M$
(dashed lines) as seen by a distant observer located on the $z$-axis.
\label{fig:gw}}


\begin{thebibliography}{}
\bibitem[Baumgarte and Shapiro(1999)]{BS} 
Baumgarte, T. W., Shapiro, S. L., 1999, 
\prd, 59, 024007
%
\bibitem[Blanchet, Damour, and Sch\"afer(1989)]{BDS}
Blanchet, L., Damour, T., and Sch\"afer, G., 1989, 
\mnras, 242, 289
%
\bibitem[Bonazzola et al.(1993)]{BGSM}
Bonazzola, S., Gourgoulhon, E.,  Salgado, M., and Marck, J. A., 1993, 
\aap, 278, 421
%
\bibitem[Bonazzola, Frieben, and Gourgoulhon(1996)]{BFG96} 
Bonazzola, S., Frieben, J., and Gourgoulhon, E.,  1996, 
\apj, 460, 379
\bibitem[Bonazzola, Frieben, and Gourgoulhon(1998)]{BFG98} 
---,  1998, \aap, 331, 280
%
\bibitem[Bowen and York(1980)]{BY}
Bowen, J. M., and York, Jr., J. W., 1980, 
\prd, 21, 2047
%
\bibitem[Brown(2000)]{Brown}
Brown, J. D., 2000, \prd, in press (gr-qc/0004002)
%
\bibitem[Carter(1979)]{Carter}
Carter,  B., 1979,
in {\em Active Galactic Nuclei}, edited by C. Hazard and S. Mitton,
(Cambridge Univ. Press, Cambridge, England), p. 273
%
\bibitem[Chandrasekhar(1965)]{Chandra65}
Chandrasekhar, S., 1965, 
\apj, 142, 1488
%
\bibitem[Chandrasekhar(1969)]{Chandra69}
---, 1969,
{\em Ellipsoidal Figures of Equilibrium},
(Yale Univ. Press, New York), p. 61
%
\bibitem[Chandrasekhar(1970)]{Chandra70}
---, 1970, \apj, 161, 561
%
\bibitem[Cook, Shapiro, and Teukolsky(1992)]{CST92} 
Cook, G. B., Shapiro S. L., and Teukolsky, S. A., 1992,
\apj, 398, 203;
\bibitem[Cook, Shapiro, and Teukolsky(1994)]{CST94} 
---, 1994, \apj, 422, 227
%
\bibitem[Cutler and Lindblom(1992)]{CL92} 
Cutler, C. and Lindblom, L., 1992, 
\apj, 385, 630
%
\bibitem[Durisen et al.(1986)]{DGTB}
Durisen, R. H., Gingold, R. A., Tohline, J. E., and Boss, A. P., 1986, 
\apj, 305, 281
%
\bibitem[Finn(1989)]{F89}
Finn, L. S., 1989, in {\it Frontiers in Numerical Relativity}, 
edited by C. R. Evans, L. S. Finn, and D. W. Hobill 
(Cambridge University Press, Cambridge, England), p. 126

\bibitem[Friedman and Schutz(1978)]{FS78} 
Friedman, J. L., and Schutz, B. F., 1978, 
\apj, 222, 281
%
\bibitem[Houser, Centrella, and Smith(1994)]{HCS}
Houser, J. L., Centrella, J. M., and Smith, S. C., 1994, 
\prl, 72, 1314
%
\bibitem[Houser and Centrella(1996)]{HC}
Houser, J. L., and Centrella, J. M., 1996, 
\prd, 54, 7278
%
\bibitem[Imamura, Friedman, and Durisen(1985)]{IFD85} 
Imamura, J. N., Friedman, J. L., and Durisen, R. H., 1985,
	\apj, 294, 474
%
\bibitem[Imamura, Toman, Durisen, Pickett, and Yang(1995)]{Imamura} 
Imamura, J. N., Toman, J., Durisen, R. H., Pickett, B. K., and Yang, S.,
1995, \apj, 444, 363
%
\bibitem[Ipser and Lindblom(1990)]{IL90}
Ipser, J. R., and Lindblom, L., 1990, 
\apj, 355, 226
%
\bibitem[Komatsu, Eriguchi, and Hachisu(1989a)]{KEH89a}
Komatsu, H., Eriguchi, Y., and Hachisu, I., 1989, 
\mnras, 237, 355 
\bibitem[Komatsu, Eriguchi, and Hachisu(1989b)]{KEH89b}
---, 1989b, \mnras, 239, 153
%
\bibitem[Landau and Lifshitz(1982)]{LL}
Landau, L. D., and Lifshitz, E. M., 1982,
{\em Fluid Mechanics, 2nd edition} (Pergamon Press, Oxford, England)
%
\bibitem[Lai and Shapiro(1995)]{LS}
Lai, D., and Shapiro, S. L., 1995, 
\apj, 442, 259
%
\bibitem[Lynden-Bell  and Ostriker(1967)]{LO67} 
Lynden-Bell, D., and Ostriker, J. P., 1967, 
\mnras, 136, 293
%
\bibitem[Misner, Thorne, and Wheeler(1973)]{MTW} 
Misner, C. W., Thorne, K. S., and Wheeler, J. A., 1973, 
{\em Gravitation} (W. H. Freeman and Company, New York),
p. 996
%
\bibitem[New, Centrella, and Tohline(2000)]{NCT}
New, K. C. B., Centrella, J. M.,  and Tohline, J. E.,  2000,
\prd, 62, 064019
%
\bibitem[Ostriker and Bodenheimer(1973)]{OB73} 
Ostriker, J. P., and Bodenheimer, P., 1973, 
\apj, 180, 171
%
\bibitem[Pickett, Durisen, and Davis(1996)]{PDD}
Pickett, B. K., Durisen, R. H., and Davis, G. A., 1996,
\apj, 458, 714
%
\bibitem[Salgado, Bonazzola, Gourgoulhon, and Haensel(1994)]{SBGH} 
Salgado, M., Bonazzola, S., Gourgoulhon, E., and Haensel, P., 1994, 
\aap, 291, 155
%
\bibitem[Shapiro and Teukolsky(1983)]{ST}
Shapiro, S. L., and Teukolsky, S. A., 1983, {Black
	Holes, White Dwarfs, and Neutron Stars}, 
(John Wiley and Sons, New York)
%
\bibitem[Shapiro and Zane(1998)]{SZ}
Shapiro, S. L., and Zane, S., 1998, 
\apjs, 117, 531
%
\bibitem[Shibata(1999a)]{Shibata99a} 
Shibata, M., 1999a, \prd, 60, 104052
%
\bibitem[Shibata(1999b)]{Shibata99b} 
---, 1999b, Prog. Theor. Phys. {101}, 1199
%
\bibitem[Shibata, Baumgarte, and Shapiro(1998)]{SBS98}
Shibata, M., Baumgarte, T. W., and Shapiro, S. L., 1998,
\prd 58, 23002
%
\bibitem[Shibata, Baumgarte, and Shapiro(2000)]{SBS00}
---, 2000, \apj, 542, 453
%
\bibitem[Shibata and Nakamura(1995)]{SN}
Shibata, M., and Nakamura, T., 1995, 
\prd, 52, 5428
%
\bibitem[Smith, Houser, and Centrella(1995)]{SHC}
Smith, S. C., Houser, J. L., and Centrella, J. M., 1995, 
\apj, 458, 236
%
\bibitem[Stergioulas and Friedman(1998)]{SF} 
Stergioulas, N., and Friedman, J. L., 1998, 
\apj, 492, 301
%
\bibitem[Tassoul(1978)]{Tassoul} 
Tassoul, J., 1978, {\em Theory of Rotating Stars} 
(Princeton University Press, Princeton, New Jersey)
%
\bibitem[Thorne(1995)]{KIP} 
Thorne, K. S., 1995, in {\em Proceeding of Snowman 
95 Summer Study on Particle and Nuclear Astrophysics and 
Cosmology}, edited by E. W. Kolb and R. Peccei (World Scientific, 
Singapore), p. 398
%
\bibitem[Tohline, Durisen, and McCollough(1985)]{TDM}
Tohline, J. E. , Durisen, R. H., and McCollough, M., 1985,
\apj, 298, 220
%
\bibitem[Tohline and Hachisu(1990)]{TH}
Tohline, J. E., and Hachisu, I., 1990, 
\apj, 361, 394 
%
\bibitem[Toman et al.(1998)]{TIPD}
Toman, J., Imamura, J. N., Pickett, B. J., and Durisen, R. H., 1998,
\apj, 497, 370
%
\bibitem[Williams and Tohline(1988)]{WT}
Williams, H. A.,  and Tohline, J. E., 1988, 
\apj, 334, 449
%
\bibitem[Yoshida and Eriguchi(1999)]{YE} 
Yoshida, S., and Eriguchi, Y., 1999, 
\apj, 515, 414
%
\end{thebibliography}
\end{document}